# From Weighted Frontier *d*-States to Geometric Descriptors for Atomically Dispersed Catalysts at the Molecule–Solid Crossover

Yuqi Li[1,†], Zewen Wu[1,†], Longling Pang[1], Zhidan Li[1], and Xianghua Kong[1,*]

**Affiliations:**

[1] College of Physics and Optoelectronic Engineering, Shenzhen University, Shenzhen 518060, China

**Abstract**

Atomically dispersed catalysts exhibit molecule–solid crossover character when localized metal orbitals hybridize with support-derived bands, a regime not fully captured by either a *d*-band-center description or a frontier-orbital picture. We establish an electronic-to-geometric framework for oxygen reduction across fourteen metals. N-to-P substitution in M–$N_4$ sites raises theoretical onset potentials in 13 systems, with Rh–$N_3$P reaching 0.88 V, comparable to the Pt(111) benchmark. Scaling relations reduce *OOH, *O, and *OH thermodynamics to $\Delta G_{*\mathrm{OH}}$. M–$N_4$ follows a $d_{yz}$-dominated frontier-orbital picture, whereas M–$N_3$P requires a weighted frontier *d*-state representation built from adsorption-relevant orbitals and their occupancies. The resulting $\varepsilon_{\mathrm{WFDS}}$ correlates with $\Delta G_{*\mathrm{OH}}$ at $R^2$ = 0.92, compared with 0.83 for the *d*-band center and 0.64 for the frontier-orbital. The identified roles of orbital occupancy and local coordination suggest an interpretable geometric descriptor, $\zeta_{\mathrm{M\text{–}N_3P}} = n_d / r_{\mathrm{M\text{–}P}}$, that predicts $\Delta G_{*\mathrm{OH}}$ at $R^2$ = 0.96. The ORR volcano analysis identifies strong-binding systems and motivates axial oxygen coordination. This design raises onset potentials by 0.70–2.46 V, and the M–$N_3$PO descriptor predicts $\Delta G_{*\mathrm{OH}}$ at $R^2$ = 0.91. By bridging frontier-orbital and *d*-band-center descriptions, the framework provides a physically grounded electronic representation and a geometry-based screening route compatible with machine-learned interatomic potentials.

*Correspondence to: kongxianghuaphysics@szu.edu.cn

†These authors contributed equally to this work.

## Introduction

Hydrogen fuel cells are a key technology for sustainable energy conversion, but their practical deployment is limited by the sluggish four-electron kinetics of the cathodic oxygen reduction reaction (ORR)[1–7]. Pt-based catalysts remain the ORR benchmark. However, their high cost and limited natural abundance motivate the development of low- or non-Pt alternatives[8–11]. Atomically dispersed catalysts (ADCs), including single-atom catalysts (SACs), dual-atom catalysts (DACs), and atomically defined metal clusters on two-dimensional supports, have attracted sustained attention, with graphene-embedded transition-metal–$N_4$ moieties (M–$N_4$) serving as a prototypical model[12–18]. These catalysts combine near-atomic-limit metal utilization with highly tunable active-site electronic structures through coordination tuning, heteroatom substitution, and axial ligand decoration[19–21]. Nevertheless, the achievable ORR activity of M–$N_4$-type catalysts remains constrained. Reported theoretical onset potentials ($U_{\mathrm{ORR}}^{\mathrm{onset}}$) for M–$N_4$ systems (M = Sc, Ti, Ni, Cu, Y, Zr, Pd, Ag, Cd, Hf, Pt, Au) range from −2.75 to 0.63 V[22], well below both the thermodynamic equilibrium potential of 1.23 V and the Pt(111) benchmark of 0.80 V[23–25]. Phosphorus substitution at N coordination sites is an attractive modulation strategy because the larger covalent radius and lower electronegativity of P relative to N can induce local structural distortion and metal-ligand charge redistribution[26,27]. This motivates us to examine whether and how such N-to-P substitution can systematically improve $U_{\mathrm{ORR}}^{\mathrm{onset}}$ in M–$N_4$-derived ADCs.

Addressing this requires moving beyond empirical screening toward a physical model that links local structure to ORR activity. ADCs anchored on two-dimensional supports belong neither to the molecular limit of discrete orbitals nor to the solid-surface limit of continuous bands. Instead, they constitute molecule–solid crossover systems, in which localized metal-centered *d* levels coexist and hybridize with delocalized states from the two-dimensional support, giving rise to a dual localized-delocalized electronic character[28–31]. In the solid-surface limit, *d*-band center theory rationalizes adsorption trends through interactions between adsorbate states and delocalized metal *d* bands[28,29]. In the molecular limit, frontier orbital theory emphasizes the energy alignment and spatial overlap between adsorbate and discrete catalyst frontier orbitals[30,31]. M–$N_4$-derived ADCs exemplify this regime[32]. Coordination-site substitution such as N-to-P further lowers the local symmetry and promotes hybridization among adsorption-relevant *d* orbitals, making the crossover character more pronounced. Consistent with this dual nature, recent studies have applied both *d*- band-based and frontier-orbital-based frameworks to ADCs[28–32]. For systems at this crossover, however, neither the $d$-band center nor a single frontier orbital is guaranteed to provide a transferable description. A single $d$-band center may obscure orbital-specific information by averaging over localized or partially hybridized $d$ states, whereas a single frontier orbital may neglect multi-state contributions that remain physically active. What is missing is an electronic-structure framework that captures both the localized and the delocalized character of these systems without collapsing onto either limit.

The third challenge concerns the computational cost of predicting activity from structure. The theoretical onset potential of ORR is typically obtained through a multi-step density functional theory (DFT) workflow, including geometry optimization, adsorption calculations for multiple intermediates, zero-point energy (ZPE) and entropy corrections, construction of free-energy diagrams, and identification of the potential-determining step[33]. Even for a minimal ORR pathway, the adsorption free energies of *OOH, *O, and *OH are required. This workload grows quickly as both the catalyst design space and the reaction-path space expand to include multiple metal centers, coordination environments, and adsorbate configurations[33]. Electronic descriptors can reduce the reliance on full reaction-path calculations, but they still require electronic-structure analyses such as projected density of states, orbital-resolved bonding, or explicit adsorbate calculations[28–31]. In this context, geometric descriptors have attracted increasing interest because they can encode catalytic trends in a more computationally economical form[34]. Bypassing costly adsorbate and electronic-structure calculations through interpretable descriptors is the methodological goal of the present work.

Here, we address these three questions through systematic first-principles calculations on M–$N_4$ and N-to-P-substituted M–$N_3P$ systems containing 14 transition metals, with axially oxidized M–$N_3PO$ sites used to test the extension to strong-binding centers. N-to-P substitution enhances $U_{\mathrm{ORR}}^{\mathrm{onset}}$ in 13 of 14 metals, with Rh–$N_3P$ reaching 0.88 V, comparable to the Pt(111) benchmark of 0.80 V. Scaling relationships among *OOH, *O, and *OH reduce ORR thermodynamics to a single $\Delta G_{*\mathrm{OH}}$ descriptor. Orbital-resolved bonding analysis further shows that M–$N_4$ follows a $d_{yz}$-dominated frontier-orbital picture, whereas M–$N_3P$ requires a weighted frontier $d$-state representation that incorporates the metal-dependent redistribution of $d$-orbital occupancies induced by N-to-P substitution. This model links multi-state hybridization, frontier-level alignment, and adsorption-relevant *d*-state filling. Guided by this picture of *d*-state filling, we construct an interpretable geometric descriptor, $\zeta_{\mathrm{M\text{–}N_3P}} = n_d / r_{\mathrm{M\text{–}P}}$, where $n_d$ captures intrinsic *d*-electron count and $r_{\mathrm{M\text{–}P}}$ encodes metal-ligand electron sharing. The descriptor predicts $\Delta G_{*\mathrm{OH}}$ for M–$N_3P$ with $R^2$ = 0.96, enabling screening without explicit adsorbate or electronic-structure calculations. Axial oxidation further improves the ORR performance of over-binding metals with fewer than eight *d* electrons, raising their onset potentials by 0.70–2.46 V, while the same geometric descriptor philosophy remains applicable. Together, this work establishes a physics-informed electronic-to-geometric descriptor framework for ADCs at the molecule–solid crossover.

## Results

### Atomic configurations and ORR activity

As illustrated in Fig. 1a, the M–$N_4$ configuration consists of a transition-metal atom embedded in a graphene divacancy and coordinated by four nitrogen atoms in a porphyrin-like geometry. Replacing one coordinating N with P breaks the local coordination symmetry and yields the M–$N_3P$ configuration. Fourteen transition metals from the 3*d*, 4*d*, and 5*d* rows (V, Cr, Nb, Mo, Tc, Ru, Rh, Pd, Ta, W, Re, Os, Ir, and Pt) are studied for both configurations.

After full geometry relaxation, the M–$N_3P$ configuration differs from M–$N_4$ in three respects. The average metal–X coordination bond length (X = N, P) in M–$N_3P$ is elongated by 0.10 to 0.16 Å (Supplementary Table 1), consistent with the larger covalent radius of P (1.10 Å) compared with N (0.71 Å). The metal atom is also displaced out of the graphene plane by −0.87 to +0.28 Å (Supplementary Table 2), with the direction depending on the metal. Bader analysis further indicates that the lower electronegativity of P (2.19) relative to N (3.04) reduces the electron transfer from the metal center to the coordinating atoms in 10 of the 14 systems by 0.11 to 0.32 *e* (Supplementary Table 3). These structural and electronic changes provide the baseline against which the activity trends are interpreted below.

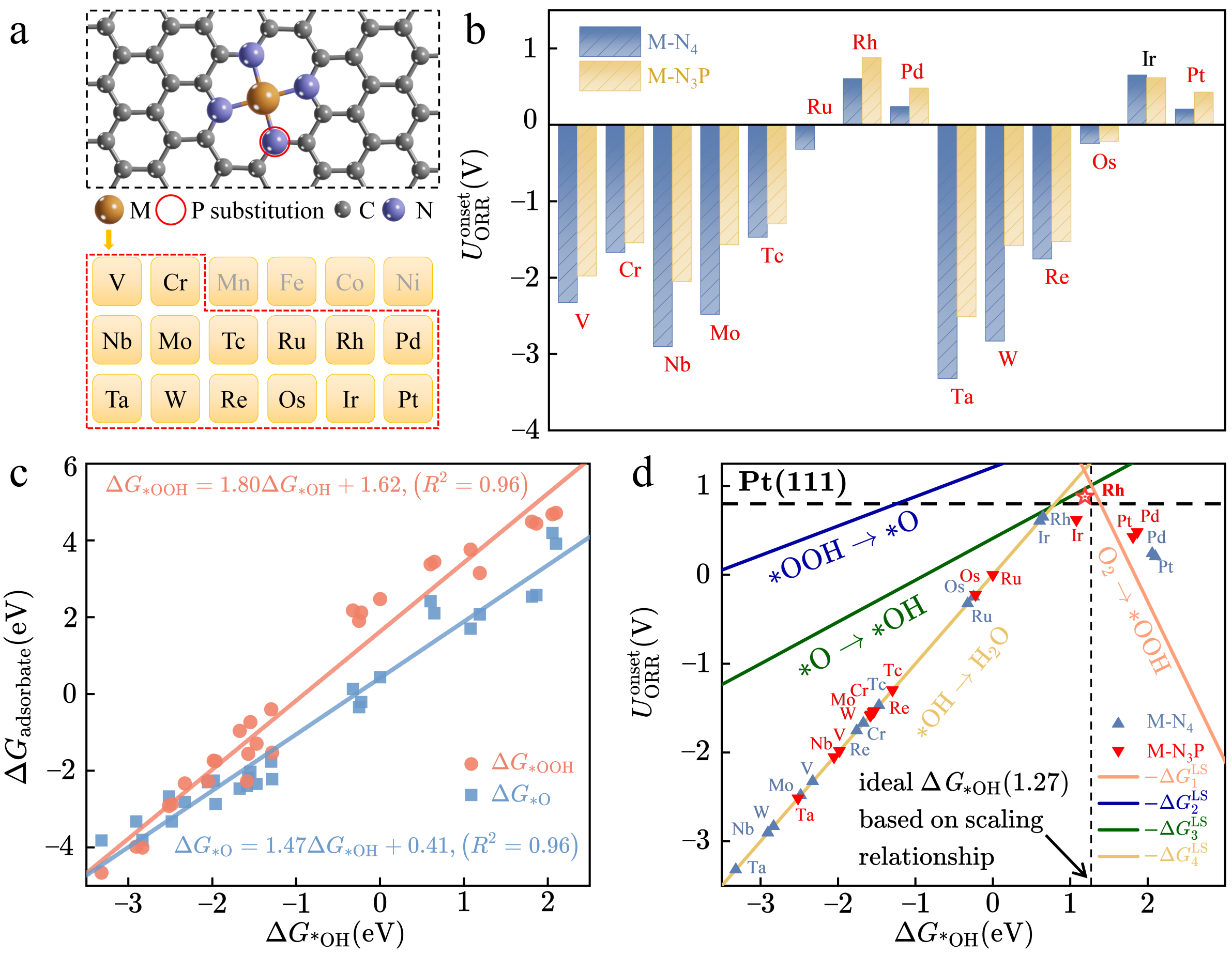

**Fig. 1 | Atomic configurations and theoretical ORR activity.** (**a**) Atomic configurations of the M–N4 and N-to-P-substituted M–N3P systems. The red circle indicates the phosphorus substitution site. (**b**) Calculated onset potentials ($U_{\mathrm{ORR}}^{\mathrm{onset}}$) for the oxygen reduction reaction (ORR) on M–N4 and M–N3P systems. (**c**) Scaling relationships among the adsorption free energies ($\Delta G_{*\mathrm{OOH}}$, $\Delta G_{*\mathrm{O}}$, and $\Delta G_{*\mathrm{OH}}$). (**d**) Sabatier volcano plot of $U_{\mathrm{ORR}}^{\mathrm{onset}}$ as a function of $\Delta G_{*\mathrm{OH}}$. The solid lines represent the negative free energy changes of the elementary steps predicted by the linear scaling relationships, and the scatter points represent the calculated onset potentials for the M–N4 and M–N3P systems. The horizontal dashed line marks the Pt(111) benchmark onset potential of 0.80 V, and the vertical dashed line indicates the ideal $\Delta G_{*\mathrm{OH}}$ of 1.27 eV.

Fig. 1b summarizes the theoretical onset potentials, $U_{\mathrm{ORR}}^{\mathrm{onset}}$, for all 28 systems. Phosphorus substitution raises $U_{\mathrm{ORR}}^{\mathrm{onset}}$ in 13 of the 14 metals studied, with gains ranging from 0.03 to 1.25 V (Supplementary Tables 4 and 5). The single exception is Ir, for which substitution lowers $U_{\mathrm{ORR}}^{\mathrm{onset}}$ by only 0.03 V. The highest activity is reached by Rh–$N_3$P at $U_{\mathrm{ORR}}^{\mathrm{onset}}$ = 0.88 V, comparable to the Pt(111) benchmark of 0.80 V. Pd–$N_3$P and Pt–$N_3$P also approach the benchmark region. These results establish N-to-P substitution as a general strategy for enhancing ORR activity in M–$N_4$ catalysts, and motivate the mechanistic analysis of the activity enhancement.

The activity trend in Fig. 1b can be traced to the adsorption strengths of the ORR intermediates. *OOH, *O, and *OH all bind to the metal center through the oxygen atom, suggesting that their adsorption free energies are governed by related M–O interactions. This is confirmed by the scaling relationships in Fig. 1c.

$$\Delta G_{*\mathrm{OOH}} = 1.80\Delta G_{*\mathrm{OH}} + 1.62, \quad R^2 = 0.96 \tag{1}$$

$$\Delta G_{*\mathrm{O}} = 1.47\Delta G_{*\mathrm{OH}} + 0.41, \quad R^2 = 0.96 \tag{2}$$

The three-dimensional ($\Delta G_{*\mathrm{OOH}}$, $\Delta G_{*\mathrm{O}}$, $\Delta G_{*\mathrm{OH}}$) space therefore reduces to a single descriptor, $\Delta G_{*\mathrm{OH}}$. Free-energy-pathway analysis identifies hydrogenation of *OH to form $H_2O$ as the potential-determining step in 22 of the 28 catalysts (Supplementary Fig. 1), supporting the use of $\Delta G_{*\mathrm{OH}}$ as the activity descriptor.

The dependence of $U_{\mathrm{ORR}}^{\mathrm{onset}}$ on $\Delta G_{*\mathrm{OH}}$ follows a Sabatier volcano (Fig. 1d). The apex sits at $\Delta G_{*\mathrm{OH}} \approx$ 1.27 eV, where the limiting free-energy changes of the ORR elementary steps are balanced under the scaling relationships. On the strong-binding branch ($\Delta G_{*\mathrm{OH}}$ < 1.27 eV), ORR is limited by *OH removal. On the weak-binding branch ($\Delta G_{*\mathrm{OH}}$ > 1.27 eV), *OOH formation becomes unfavourable. N-to-P substitution shifts catalysts on both branches toward the apex. It weakens excessive M–OH binding for strong-binding metals and strengthens insufficient M–OH binding for weak-binding metals. Phosphorus substitution therefore acts as a bidirectional regulator of M–OH binding rather than inducing a uniform shift in adsorption strength, accounting for the activity gains in 13 of 14 metals and for the benchmark-comparable performance of Rh–$N_3$P. The direction of this shift in a given system reflects the metal's local electronic structure, which we examine in the next section.

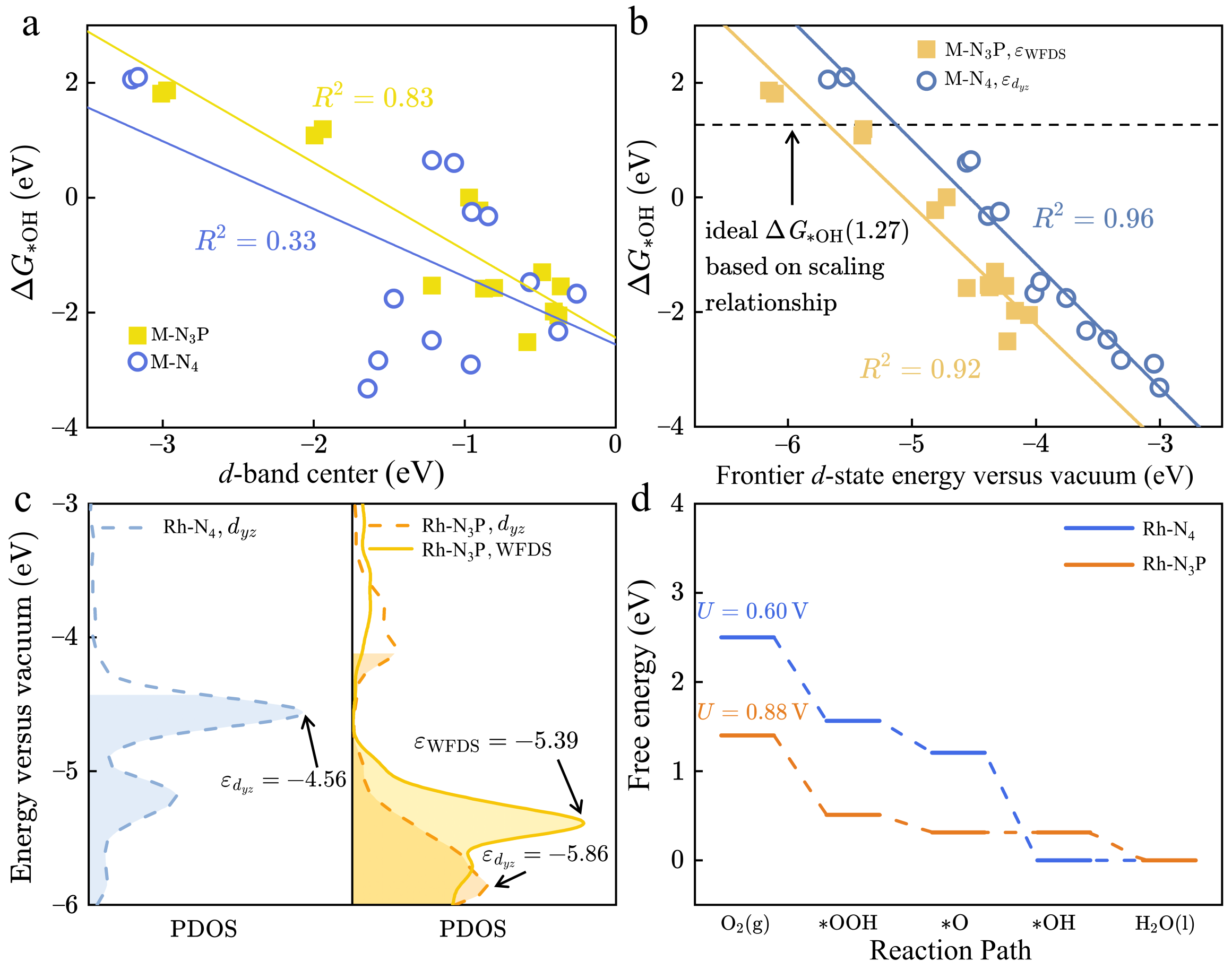

**Fig. 2 | Electronic-structure origins of the ORR activity and corresponding free-energy diagrams.** (**a**) Correlation between the *d*-band center and $\Delta G_{*\mathrm{OH}}$ for the M–N4 and M–N3P systems. (**b**) Linear dependence of $\Delta G_{*\mathrm{OH}}$ on $\varepsilon_{d_{yz}}$ for M–N4 and on the weighted frontier *d*-state position $\varepsilon_{\mathrm{WFDS}}$ for M–N3P. (**c**) PDOS for Rh–N4 and Rh–N3P. (**d**) Free-energy diagrams for the ORR pathway on Rh–N4 and Rh–N3P at their respective $U_{\mathrm{ORR}}^{\mathrm{onset}}$.

## Weighted frontier *d*-state representation

We first test the *d*-band center against $\Delta G_{*\mathrm{OH}}$ for the 28 M–N4 and M–N3P systems (Fig. 2a). The correlation is weak for M–N4 ($R^2$ = 0.33) and moderate for M–N3P ($R^2$ = 0.83), indicating that the *d*-band center is insufficient to describe $\Delta G_{*\mathrm{OH}}$ in either series. This limitation is consistent with their distinct partial density of states (PDOS) characteristics. In M–N4, the metal *d* orbitals retain localized, atom-like discrete energy levels, with relatively separated orbital features (Supplementary Fig. 2). In M–$N_3$P, N-to-P substitution lowers the local symmetry and induces inter-orbital hybridization, giving rise to broader and more band-like *d*-state features (Supplementary Fig. 2). Therefore, although the *d*-band center partially captures the multi-state character of M–$N_3$P, it does not resolve the orbital-specific information needed to describe M–OH bonding. This motivates an orbital-resolved analysis of the electronic origin of $\Delta G_{*\mathrm{OH}}$.

For M–$N_4$, the localized electronic structure permits a frontier-orbital description of *OH adsorption. We quantify the contribution of each metal *d* orbital to the M–OH bond using integrated crystal orbital Hamilton population (ICOHP) analysis[35]. As shown in Supplementary Fig. 3a, the $d_{yz}$ orbital gives the strongest correlation between its ICOHP values and $\Delta G_{*\mathrm{OH}}$ ($R^2$ = 0.94), with $d_{xz}$ also showing a high but slightly weaker correlation ($R^2$ = 0.90). The PDOS analysis further shows that $d_{yz}$ exhibits a localized and well-defined frontier-like peak (Supplementary Fig. 2), making its peak position a suitable orbital-alignment descriptor. Using the $d_{yz}$ peak position $\varepsilon_{d_{yz}}$ as the descriptor gives a strong linear correlation with $\Delta G_{*\mathrm{OH}}$ ($R^2$ = 0.96, Fig. 2b), demonstrating that M–$N_4$ is well described by a $d_{yz}$-dominated frontier-orbital picture.

With respect to M–$N_3$P, however, ICOHP analysis no longer identifies a single dominant *d* orbital. The $d_{yz}$, $d_{xz}$, and $d_{z^2}$ orbitals show comparable ICOHP correlations with $\Delta G_{*\mathrm{OH}}$, with $R^2$ values in the range of 0.70–0.73 (Supplementary Fig. 3b), indicating that M–OH bonding is distributed over multiple $d$-orbital components. This behavior is consistent with the broader and hybridized PDOS features of M–$N_3$P (Supplementary Fig. 2). Moreover, $\Delta G_{*\mathrm{OH}}$ exhibits a strong linear correlation with the total metal *d*-orbital occupancy, with $R^2$ = 0.97 for M–$N_3$P and $R^2$ = 0.95 for M–$N_4$ (Supplementary Fig. 4). This trend indicates that the metal *d*-state occupancy is a key electronic factor governing M–OH binding. The orbital-resolved occupancy map in Supplementary Fig. 5 further shows that the redistribution of metal *d*-orbital occupancies induced by N-to-P substitution is metal-dependent. Given that *OH binds to the metal center through its oxygen atom, this filling response provides a physical basis for resolving the *d*-orbital contributions to M–O bonding. Accordingly, the weighted frontier *d*-state PDOS is constructed from a metal-specific orbital set, $\Omega_{\mathrm{M}}$, comprising the *d* orbitals whose calculated occupancies increase upon substitution, as summarized in Supplementary Table 6:

$$\rho_{\mathrm{WFDS}}(E) = \sum_{i \in \Omega_{\mathrm{M}}} n_i \rho_i(E) \quad (3)$$

Here, $\rho_i(E)$ and $n_i$ denote the orbital-resolved PDOS and electron occupancy of orbital *i* in M–$N_3P$, respectively. The substitution-induced occupancy change determines the adsorption-relevant orbital set $\Omega_{\mathrm{M}}$, whereas the absolute occupancy in M–$N_3P$ sets the weight of each component to the summed PDOS. The frontier peak position of $\rho_{\mathrm{WFDS}}(E)$, denoted $\varepsilon_{\mathrm{WFDS}}$, is taken as the weighted frontier *d*-state energy position.

The resulting $\varepsilon_{\mathrm{WFDS}}$ is distinct from the *d*-band center. It is not obtained by averaging over all five metal *d* orbitals or over the full *d*-state distribution, but is extracted from a weighted PDOS constructed from the adsorption-relevant *d*-orbital components. As shown in Fig. 2b, $\varepsilon_{\mathrm{WFDS}}$ correlates with $\Delta G_{*\mathrm{OH}}$ for M–$N_3P$ with $R^2 = 0.92$, compared with $R^2 = 0.83$ for the *d*-band center in Fig. 2a and $R^2 = 0.64$ for the single-frontier-orbital description in Supplementary Fig. 6. The weighted frontier *d*-state representation establishes a multi-orbital frontier-state picture for M–$N_3P$ and bridges the two limits of the molecule–solid crossover. It incorporates the multi-orbital character of *d*-band-based descriptions while preserving the energy-alignment principle of frontier-orbital theory through the frontier peak position of $\rho_{\mathrm{WFDS}}(E)$.

Fig. 2c illustrates the transition from the $d_{yz}$-dominated frontier-orbital picture in M–$N_4$ to the weighted frontier *d*-state representation in M–$N_3P$, using Rh as a representative case. In Rh–$N_4$, the $d_{yz}$ frontier peak is located at −4.56 eV relative to vacuum and provides a direct orbital reference for comparing Rh–$N_4$ and Rh–$N_3P$. After N-to-P substitution, the $d_{yz}$-related feature shifts downward to −5.86 eV, indicating that the original $d_{yz}$-mediated Rh–OH interaction is weakened under the same orbital reference. This single-orbital comparison identifies the direction of the electronic modulation, whereas $\varepsilon_{\mathrm{WFDS}}$ provides the quantitative frontier-energy measure for the multi-orbital M–$N_3P$ system. In Rh–$N_3P$, P-induced *d*-orbital hybridization is accompanied by a redistribution of the orbital occupancies. The metal-specific set $\Omega_{\mathrm{Rh}}$ therefore comprises $d_{xz}$ and $d_{x^2-y^2}$, whose calculated occupancies increase upon substitution (Supplementary Table 6). The corresponding occupancy-weighted PDOS exhibits a frontier peak at −5.39 eV, which defines $\varepsilon_{\mathrm{WFDS}}$. Using the appropriate frontier representation for each coordination environment, the characteristic frontier energy is 0.83 eV lower in Rh–$N_3P$ than in Rh–$N_4$ (−5.39 versus −4.56 eV). This downshift increases the energy separation from the O-centered acceptor states of *OH, weakens the Rh–OH interaction, and moves $\Delta G_{*\mathrm{OH}}$ toward the volcano apex. The free-energy diagrams in Fig. 2d show that $U_{\mathrm{ORR}}^{\mathrm{onset}}$ increases accordingly, from 0.60 V for Rh–$N_4$ to 0.88 V for Rh–$N_3$P.

**Geometric descriptors for activity prediction**

The weighted frontier *d*-state picture indicates that M–OH binding is governed by both metal identity and relaxed local coordination geometry. The metal center provides the intrinsic *d*-electron count, whereas the relaxed coordination environment determines how this electronic population is redistributed through metal–ligand bonding. For the M–$N_3$P series, the metal identity is represented by the metal *d*-electron count $n_d$, and the primary coordination perturbation introduced by N-to-P substitution is represented by the relaxed M–P bond length $r_{\mathrm{M\text{–}P}}$. We therefore define an interpretable geometric descriptor,

$$\zeta_{\mathrm{M\text{–}N_3P}} = n_d / r_{\mathrm{M\text{–}P}} \tag{4}$$

As shown in Fig. 3a, $\zeta_{\mathrm{M\text{–}N_3P}}$ exhibits a strong linear relationship with $\Delta G_{*\mathrm{OH}}$,

$$\Delta G_{*\mathrm{OH}} = 1.76\ \zeta_{\mathrm{M\text{–}N_3P}} - 5.42, \quad R^2 = 0.96 \tag{5}$$

The positive slope is consistent with the *d*-state occupancy picture established above. Increasing $\zeta_{\mathrm{M\text{–}N_3P}}$, through a larger metal *d*-electron count or a shorter M–P bond length, corresponds to weaker M–OH binding and a higher $\Delta G_{*\mathrm{OH}}$. The M–$N_4$ series follows the same geometry-based trend with $\zeta_{\mathrm{M\text{–}N_4}} = n_d / \bar{r}_{\mathrm{M\text{–}N}}$, where $\bar{r}_{\mathrm{M\text{–}N}}$ is the average M–N bond length ($R^2$ = 0.96 in Supplementary Fig. 7). This result shows that the dependence of OH adsorption thermodynamics on metal *d*-electron count and local bond length also holds for the parent N-coordinated environment.

The volcano relation established in Fig. 1d links $\Delta G_{*\mathrm{OH}}$ to $U_{\mathrm{ORR}}^{\mathrm{onset}}$. Combining this relation with the $\zeta_{\mathrm{M\text{–}N_3P}}$–$\Delta G_{*\mathrm{OH}}$ correlation in Fig. 3a gives the structure–activity relationship shown in Fig. 3b. M–$N_3$P systems with $n_d$ < 8 populate the strong-binding branch, where excessive M–OH binding suppresses the onset potential. Rh–$N_3$P and Ir–$N_3$P lie near the volcano apex, corresponding to nearly optimal M–OH binding. Pd–$N_3$P and Pt–$N_3$P approach the weak-binding branch, where *OOH formation becomes less favorable. Thus, $\zeta_{\mathrm{M\text{–}N_3P}}$ links metal *d*-electron count and relaxed M–P geometry to both $\Delta G_{*\mathrm{OH}}$ and $U_{\mathrm{ORR}}^{\mathrm{onset}}$, and distinguishes the binding regimes that determine the direction of further optimization.

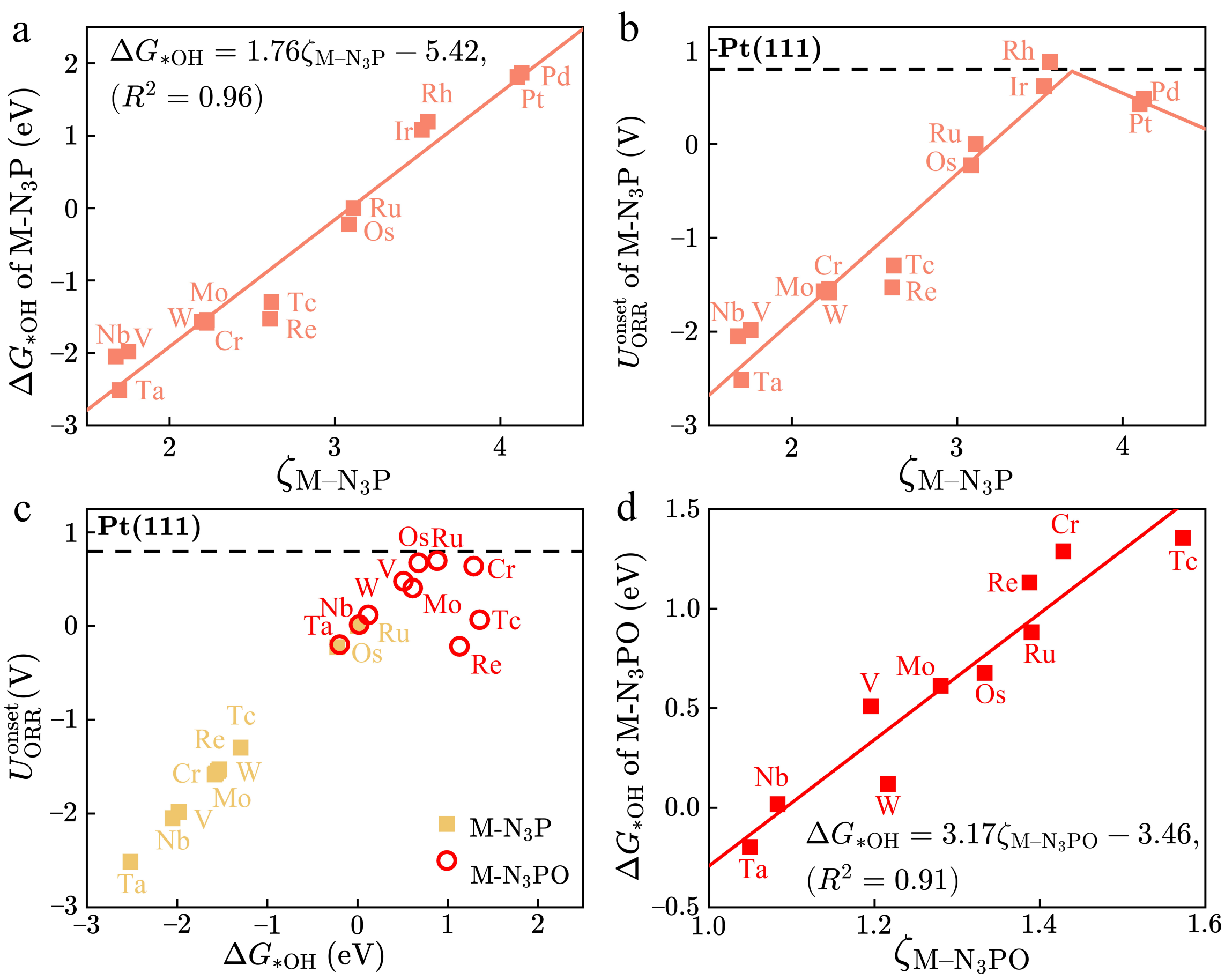

**Fig. 3 | Structural descriptors and ORR activity regulation via axial oxidation.** (**a**) Adsorption free energy ($\Delta G_{*\mathrm{OH}}$) as a function of the structural descriptor $\zeta_{\mathrm{M\text{-}N_3P}}$. (**b**) Corresponding dependence of the theoretical onset potential ($U_{\mathrm{ORR}}^{\mathrm{onset}}$) on the structural descriptor $\zeta_{\mathrm{M\text{-}N_3P}}$. (**c**) Volcano plot comparison for M–N3P systems with $n_d < 8$ before and after axial oxidation (to form M–N3PO). (**d**) Relationship between the adsorption free energy ($\Delta G_{*\mathrm{OH}}$) and the structural descriptor $\zeta_{\mathrm{M\text{-}N_3PO}}$ for the M–N3PO systems.

## Activity regulation via axial oxidation

The structure–activity relationship in Fig. 3b further identifies the systems that require additional structural regulation. The ten $n_d$ < 8 metals, namely V, Cr, Nb, Mo, Tc, Ru, Ta, W, Re, and Os, remain on the strong-binding branch after N-to-P substitution. For these systems, in-plane coordination tuning is insufficient to alleviate excessive M–OH binding. Guided by the *d*-state occupancy and local-coordination picture established above, we introduce an axial oxygen ligand as an additional coordination perturbation to further redistribute the metal *d*-state population and weaken the residual M–OH interaction. The resulting M–$N_3$PO structures therefore test whether the same physical picture can guide optimization beyond in-plane substitution. As shown in Fig. 3c, axial oxidation shifts all ten systems toward weaker M–OH binding and higher onset potentials. The $U_{\mathrm{ORR}}^{\mathrm{onset}}$ values increase by 0.70–2.46 V, with eight of the ten systems entering the positive-potential region (Supplementary Tables 5 and 7). Notably, Ru–$N_3$PO reaches 0.70 V, Os–$N_3$PO reaches 0.68 V, and Cr–$N_3$PO reaches 0.64 V, approaching the Pt(111) benchmark of 0.80 V.

We next examine whether this geometry-based description remains valid after introducing the axial oxygen ligand. For M–$N_3$PO, the local coordination environment contains both the axial M–O bond and the in-plane M–N/P coordination scaffold. We therefore define

$$\zeta_{\mathrm{M\text{–}N_3PO}} = (n_d/r_{\mathrm{M\text{–}O}})(\bar{r}_{\mathrm{M\text{–}N,P}} - r_{\mathrm{M\text{–}O}}) \tag{6}$$

where $r_{\mathrm{M\text{–}O}}$ is the axial metal–oxygen bond length and $\bar{r}_{\mathrm{M\text{–}N,P}}$ is the mean in-plane metal–ligand bond length. This form can be written as $n_d(\bar{r}_{\mathrm{M\text{–}N,P}}/r_{\mathrm{M\text{–}O}} - 1)$, making the axial M–O distance an explicit structural variable. The sensitivity of $\zeta_{\mathrm{M\text{–}N_3PO}}$ to $r_{\mathrm{M\text{–}O}}$ is larger than its sensitivity to $\bar{r}_{\mathrm{M\text{–}N,P}}$, because $\bar{r}_{\mathrm{M\text{–}N,P}}/r_{\mathrm{M\text{–}O}} > 1$ for the optimized M–$N_3$PO structures considered here (Supplementary Table 8). This feature is consistent with the design rationale that axial oxygen provides the additional coordination perturbation in M–$N_3$PO. Fig. 3d shows that $\zeta_{\mathrm{M\text{–}N_3PO}}$ retains a strong linear relationship with $\Delta G_{*\mathrm{OH}}$,

$$\Delta G_{*\mathrm{OH}} = 3.17\zeta_{\mathrm{M\text{–}N_3PO}} - 3.46, \quad R^2 = 0.91 \tag{7}$$

This result shows that the electronic-to-geometric relationship established for M–$N_3$P remains predictive after axial oxidation.

The progression from the scaling relationships to $\varepsilon_{\mathrm{WFDS}}$ and finally to $\zeta_{\mathrm{M\text{–}N_3P}}$ and $\zeta_{\mathrm{M\text{–}N_3PO}}$ establishes an electronic-to-geometric framework for activity prediction. The linear scaling relationships among *OOH, *O, and *OH first reduce the multi-intermediate ORR free-energy description to a single thermodynamic variable, $\Delta G_{*\mathrm{OH}}$ . The weighted frontier *d*-state representation then links $\Delta G_{*\mathrm{OH}}$ to the adsorption-relevant electronic structure. The geometric descriptors further encode this electronic information using metal identity and relaxed local bond lengths. This hierarchy enables ORR activity trends to be predicted from metal identity and relaxed local geometry while avoiding explicit adsorption calculations and orbital-resolved electronic-structure analysis during screening.

**Discussion**

In this work, we systematically investigated $M–N_4$, N-to-P-substituted $M–N_3P$, and axially oxidized $M–N_3PO$ atomically dispersed catalysts for the oxygen reduction reaction. Across fourteen transition-metal centers spanning the 3*d*, 4*d*, and 5*d* rows, N-to-P substitution enhances the theoretical onset potential in 13 of the 14 metals. $Rh–N_3P$ reaches $U_{\mathrm{ORR}}^{\mathrm{onset}}$ = 0.88 V, comparable to the Pt(111) benchmark of 0.80 V. Linear scaling relationships among *OOH, *O, and *OH reduce the multi-intermediate ORR free-energy description to a single thermodynamic variable, $\Delta G_{*\mathrm{OH}}$, establishing M–OH binding as the central quantity for interpreting the activity changes induced by phosphorus substitution.

The electronic origin of this activity regulation reflects the molecule–solid crossover character of the active site. $M–N_4$ retains localized metal *d* states and follows a $d_{yz}$-dominated frontier-orbital picture. In contrast, $M–N_3P$ exhibits stronger orbital hybridization, and its M–OH bonding is distributed over multiple *d*-orbital components. To describe this hybridized regime, we establish a weighted frontier *d*-state representation by constructing an occupancy-weighted PDOS from metal-specific *d*-orbital components identified by their P-induced occupancy increases and extracting the frontier peak position, $\varepsilon_{\mathrm{WFDS}}$. This energy position correlates with $\Delta G_{*\mathrm{OH}}$ for $M–N_3P$ with $R^2$ = 0.92, improving over the *d*-band-center correlation of $R^2$ = 0.83. The weighted frontier *d*-state representation incorporates the multi-orbital character of *d*-band-based descriptions while preserving the energy-alignment principle of frontier orbital theory.

Guided by this electronic picture, we further encode the key electronic information in interpretable geometric descriptors. For $M–N_3P$, $\zeta_{\mathrm{M\text{–}N_3P}} = n_d/r_{\mathrm{M\text{–}P}}$ predicts $\Delta G_{*\mathrm{OH}}$ with $R^2$ =

0.96, linking the metal $d$-electron count and relaxed M–P bond length to OH adsorption thermodynamics. For strong-binding metals with $n_d < 8$, axial oxidation to M–$N_3$PO further increases $U_{\mathrm{ORR}}^{\mathrm{onset}}$ by 0.70–2.46 V. The corresponding descriptor $\zeta_{\mathrm{M\text{–}N_3PO}}$, which incorporates axial M–O bonding and in-plane coordination imbalance, remains predictive with $R^2$ = 0.91. These results establish an electronic-to-geometric framework for atomically dispersed catalysts at the molecule–solid crossover, enabling ORR activity trends to be predicted from metal identity and relaxed coordination geometry while reducing reliance on explicit adsorption calculations and orbital-resolved electronic-structure analysis.

Beyond this specific M–$N_4$-derived platform, the underlying principle should be relevant to catalytic systems in which localized frontier orbitals coexist and hybridize with extended band-like states. In such molecule–solid crossover systems, weighted frontier-state representations can serve as a bridge between $d$-band-based and frontier-orbital-based models. When combined with machine-learning force fields that enable efficient geometry optimization, system-specific geometric descriptors may further accelerate structure-based catalyst screening beyond explicit DFT adsorption workflows.

## Methods

### DFT Calculation Details

DFT calculations were performed using the Vienna Ab initio Simulation Package (VASP 6.2.1)[36,37]. The electron-ion interactions were described by the projector augmented wave (PAW) method[38]. The generalized gradient approximation (GGA) with the optB86b-vdW functional was employed to treat the exchange-correlation energy, ensuring an accurate description of the van der Waals (vdW) interactions[39,40]. The plane-wave energy cutoff was set to 500 eV for geometry optimizations and increased to 600 eV for self-consistent field (SCF) and electronic property calculations to ensure high precision.

The atomically dispersed catalyst models (M–$N_4$, M–$N_3$P, and M–$N_3$PO) were constructed based on a 9 × 9 × 1 graphene supercell. A vacuum layer of 35 Å was applied along the $z$-direction to prevent spurious interactions between periodic images. The Brillouin zone was sampled using a Γ-centered 3 × 3 × 1 $k$-point grid. Geometry optimizations were performed until the residual forces on each atom were less than 0.02 eV $\text{Å}^{-1}$. For the accurate determination of electronic structures, the tetrahedron method with Blöchl corrections was adopted.

**Free Energy Calculations**

The oxygen reduction reaction (ORR) in acidic media proceeds via a four-electron pathway involving *OOH, *O, and *OH intermediates. The elementary steps are as follows:

$$\begin{gathered} \mathrm{O_2(g) + H^+ + e^- + * \rightarrow {}^*OOH} \\ \mathrm{{}^*OOH + H^+ + e^- \rightarrow {}^*O + H_2O(l)} \\ \mathrm{{}^*O + H^+ + e^- \rightarrow {}^*OH} \\ \mathrm{{}^*OH + H^+ + e^- \rightarrow * + H_2O(l)} \end{gathered} \tag{8}$$

where * denotes the active site on the substrate, and (g) and (l) refer to the gas and liquid phases, respectively.

To obtain the adsorption free energies of the reaction intermediates, vibrational frequency calculations were performed using the finite displacement method with a step size of 0.005 Å. Only the adsorbates were allowed to vibrate, while the substrate atoms were fixed. The zero-point energy (ZPE) and entropy ($S$) terms were processed and extracted using the VASPKIT code[41,42]. The Gibbs free energy change ($\Delta G$) for each elementary step was calculated based on the computational hydrogen electrode (CHE) model[25]:

$$\Delta G = \Delta E + \Delta E_{\mathrm{ZPE}} - T\Delta S + \Delta G_U \tag{9}$$

where $\Delta E$ is the reaction energy difference obtained from DFT calculations; $\Delta E_{\mathrm{ZPE}}$ and $\Delta S$ are the changes in zero-point energy and entropy, respectively. $\Delta G_U = -neU$ accounts for the effect of the applied electrode potential $U$.

The Gibbs free energy changes for the four elementary steps ($\Delta G_1$ to $\Delta G_4$) can be expressed as:

$$\begin{gathered} \Delta G_1 = \Delta G_{*\mathrm{OOH}} - 4.92 \\ \Delta G_2 = \Delta G_{*\mathrm{O}} - \Delta G_{*\mathrm{OOH}} \\ \Delta G_3 = \Delta G_{*\mathrm{OH}} - \Delta G_{*\mathrm{O}} \\ \Delta G_4 = -\Delta G_{*\mathrm{OH}} \end{gathered} \tag{10}$$

The theoretical onset potential ($U_{\mathrm{ORR}}^{\mathrm{onset}}$) is calculated as:

$$U_{\mathrm{ORR}}^{\mathrm{onset}} = -\max[\Delta G_1, \Delta G_2, \Delta G_3, \Delta G_4]/e \tag{11}$$

**Data Availability:** All data that support the findings of this study are present in the paper and the Supplementary Information. Further information can be acquired from the corresponding authors upon reasonable request.

**Acknowledgements**

We gratefully acknowledge the financial support from the Shenzhen Science and Technology Innovation Commission under the Outstanding Youth Project (Grant No. RCYX20231211090126026), the National Natural Science Foundation of China (Grants No. 12474173, 52461160327), and Department of Science and Technology of Guangdong Province (Grants No. 2021QN02L820). The computational work was supported by the in-house computing cluster SPACE and the computational facilities provided by HZWTECH.

**Author contributions**

X. K. conceived and supervised the project. Y. L. and Z. W. performed the DFT calculations. X. K., Y. L. and Z. W. analysed the results. Y. L. prepared the figures with guidance from X. K., and L. P. and Z. L. assisted in the figure preparation. Y. L. and X. K. wrote the manuscript. All authors discussed the results and revised the manuscript.

**Competing interests**

The authors declare no competing interests.